\documentclass[12pt,a4paper]{article}

\makeatletter 
\def\set@curr@file#1{%
  \begingroup
    \escapechar\m@ne
    \xdef\@curr@file{\expandafter\string\csname #1\endcsname}%
  \endgroup
}
\def\quote@name#1{"\quote@@name#1\@gobble""}
\def\quote@@name#1"{#1\quote@@name}
\def\unquote@name#1{\quote@@name#1\@gobble"}
\makeatother

\usepackage{graphicx}
\usepackage{xcolor}
\usepackage[normalem]{ulem}
\usepackage{url}
\usepackage{amssymb}
\usepackage{ulem}
\usepackage{authblk}
\usepackage[left=2.25cm, right=2.25cm, top=2cm, bottom=2.5cm]{geometry}

\date{}

\title{\textbf{Strain-engineered wrinkles on graphene using polymeric actuators}}

\author{
Davide Giambastiani$^{1,2}$, 
Cosimo Tommasi$^{1,2}$, 
Federica Bianco$^{2}$, Filippo Fabbri$^{2}$, Camilla Coletti$^{3,4}$,
Alessandro Tredicucci$^{1,2}$,
Alessandro Pitanti$^{2}$ and 
Stefano Roddaro$^{1,2}$
}

\affil{$^1$ Dipartimento di Fisica "E. Fermi", Università di Pisa, Largo B. Pontecorvo 3, I-56127 Pisa, Italy}
\affil{$^2$ NEST, CNR - Istituto Nanoscienze and Scuola Normale Superiore, Piazza San Silvestro 12, I-56127 Pisa, Italy}
\affil{$^3$ Center for Nanotechnology Innovation @NEST, Istituto Italiano di Tecnologia, Piazza San Silvestro 12, I-56127 Pisa, Italy}
\affil{$^4$ Graphene Labs, Istituto Italiano di Tecnologia, Via Morego 30, I-16163 Genova, Italy}

\begin{document}

\maketitle

\begin{abstract}
The electronic and optical properties of graphene can be precisely tuned by generating deterministic arrangements of strain features. In this paper, we report the formation of widespread and controlled buckling delamination of monolayer graphene deposited on hexagonal boron-nitride promoted by a significant squeezing of the graphene flake and induced by polymeric micro-actuators. The flexibility of this method offers a promising technique to create arbitrary buckling geometries and arrays of wrinkles which could also be subjected to iterative folding-unfolding cycles. Further development of this method could pave the way to tune the properties of several kinds of other two-dimensional materials, such as transition metal dichalcogenides, by tailoring their surface topography.
\end{abstract}

\section{Introduction}

Morphological corrugations are often undesirable structures in electronic devices since they can contribute, for example, to electrical performance degradation and transport anisotropy in graphene \cite{zhu2012structure}. However, the presence of wrinkles and ripples can be exploited to tune the electronic, optical, chemical and mechanical properties of graphene and other two-dimensional materials \cite{deng2016wrinkled,chen2018wrinkling}. In fact, these structures can be used to induce electron confinement by opening a band gap \cite{lim2015structurally} or strongly localized pseudo-magnetic fields \cite{yan2012observation,meng2013strain,ma2018landau} leading to pseudo-Landau quantization of the energy bands. It has been shown that electric conduction across wrinkles can induce the Peltier effect giving rise to temperature gradients between the two sides of a single wrinkle \cite{hu2020enhanced}. Moreover, wrinkles have also been employed to realize ultrathin pressure sensors \cite{chen2017structural} and stretchable high performance supercapacitors for energy storage applications \cite{chen2014transparent}. The modulation of wrinkle structures could also be advantageous to realize valley-polarized channels \cite{carrillo2016strained}, for controllable hydrogen adsorption storage \cite{goler2013influence} and wearable electronic devices \cite{chen2018wrinkling}.

Wrinkles can form accidentally, during device fabrication or the growth of graphene \cite{deng2016wrinkled,chen2018wrinkling}; moreover, gas trapped between graphene and its supporting substrate can induce corrugations associated to blisters of several shapes and sizes \cite{stolyarova2009observation,georgiou2011graphene,pan2012biaxial}. Along with accidental generation, the controlled creation of corrugations would be highly desirable. A number of techniques have been developed to this end: thermal contraction/expansion of the substrate \cite{Bao2009} and optical excitation \cite{hu2016rippling} have been used for periodic patterns of ripples. Pre-stressed polymeric materials \cite{zang2013,lee2016} or substrates with ordered micro-structures \cite{Reserbat2014} have been used to create corrugations on graphene. All these methods are very successful but they require a fixed patterned substrate geometry (which cannot be further modified) or the fabrication of fragile architectures, such as  suspended membranes.

On the contrary, recently developed polymeric actuators based on poly-methyl-methacrylate (PMMA) provide a simple and flexible method to strain \cite{colangelo2018controlling} and manipulate \cite{gilbert2019strain} graphene. In appropriate configurations, they can be exploited to tune the strain and morphology of graphene by creating complex and controllable networks of wrinkles. This kind of actuators can also be subjected to multiple pulling/relaxation cycles \cite{colangelo2018controlling} in order to iteratively fold and unfold individual wrinkles, paving the way to tunable and reconfigurable devices \cite{li2015surface,kang2018mechanically}.
\begin{figure}
    \centering
    \includegraphics[scale=0.35]{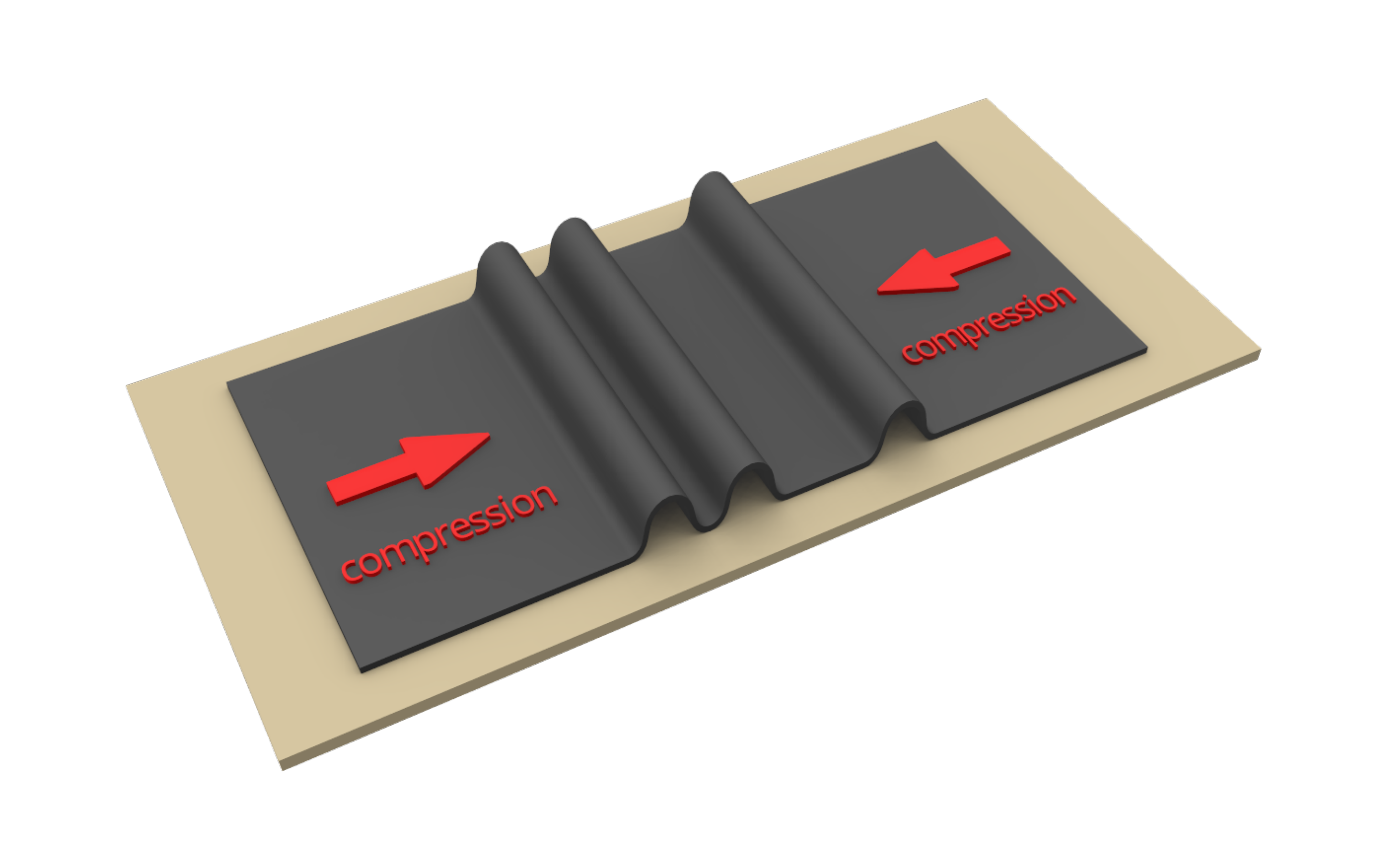}
    \caption{Schematic representation of the localized wrinkle formation that is expected to emerge when a two-dimensional surface is subjected to an in-plane
compressive strain.}
    \label{fig:intro}
\end{figure}
In this paper, we exploit the low-friction between graphene and hexagonal boron-nitride (hBN) in a van der Waals heterostructure \cite{leven2013robust,wang2016thermally} to induce squeezing of graphene by means of polymeric actuators patterned on top of the flake. This promotes the formation of a controllable wrinkle pattern (see Fig.~\ref{fig:intro}), whose arrangement depends on both the geometry of the graphene edges, and the geometry of the polymeric actuators used.

In section \ref{sec:characterization}, we show the characterization of a typical graphene flake via Atomic Force microscopy and Raman measurements. In section \ref{sec:crumpling}, we demonstrate the creation of a complex and reproducible network of wrinkles. In section \ref{sec:sim}, we report on numerical simulations and motivate the observed wrinkle pattern geometry. In section \ref{sec:conclusion}, we discuss the results and future perspectives.

\section{Characterization of graphene} \label{sec:characterization}
\begin{figure*}[h!]
    \centering
    \includegraphics[scale=0.25]{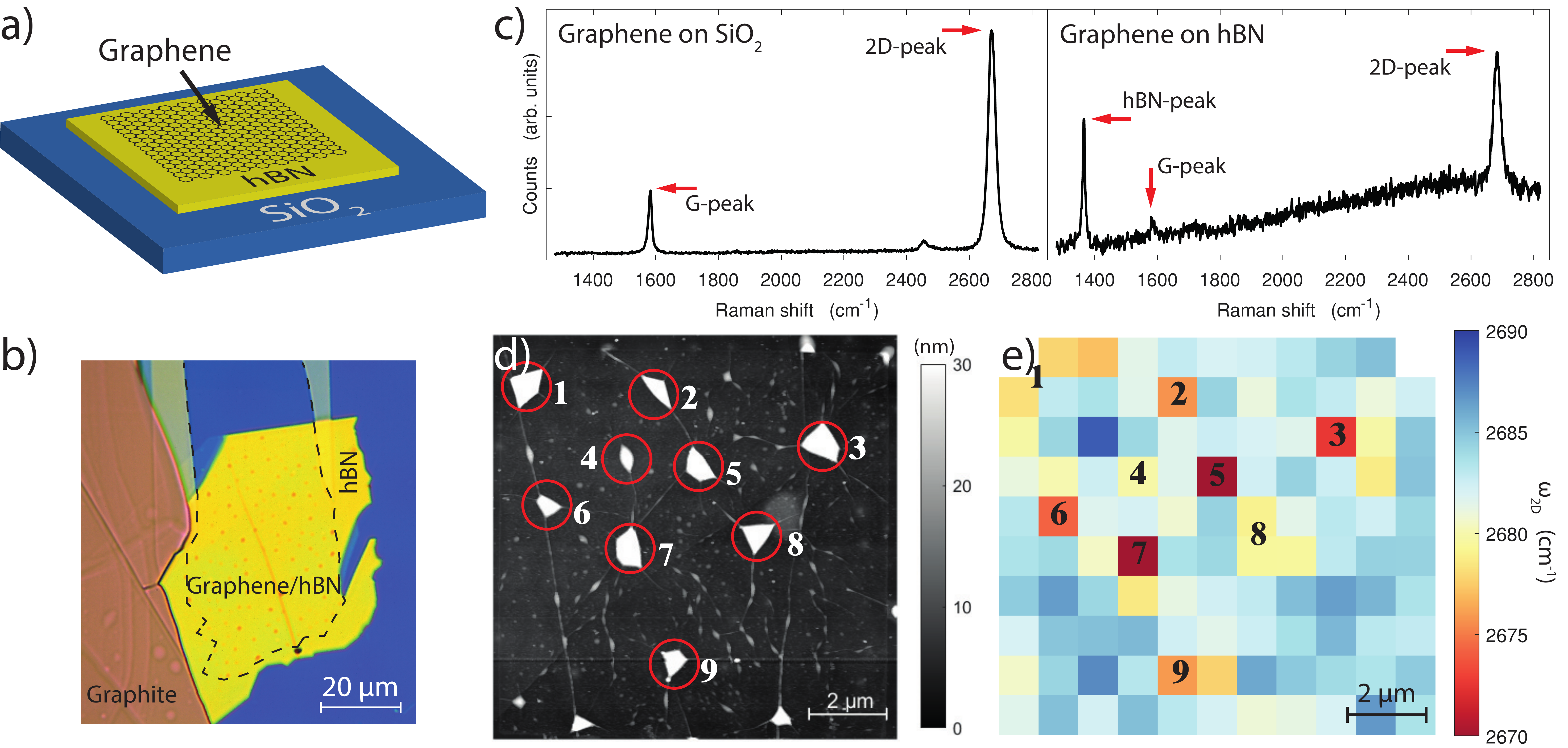}
    \caption{a) Schematic sketch of the transferred graphene stacked on hBN. b) Optical micrograph of graphene transferred on hBN. The dashed line follows the edge of the graphene flake. c) Typical Raman spectra of the initial pre-etched graphene deposited on SiO$_2$ (left) and on hBN (right). d) AFM scan of the 10$\times$10 $\mu$m$^2$ graphene flake. e) Raman map of the 2D peak position of the etched flake displayed in panel (d).}
    \label{fig:characterization}
\end{figure*}
In order to fabricate the graphene/hBN heterostructures, we used a polymer-assisted transfer technique, as described in Appendix \ref{app:fab}. We identified some flakes of graphene exfoliated on PMMA and we transferred them targeting selected hBN flakes, previously exfoliated onto a 300 nm SiO$_2$ layer. We selected only thick hBN flakes (40-80 nm) in order to have small hBN roughness \cite{dean2010boron} and therefore small frictional forces at the graphene/hBN interface, leading to enhanced slipping and easier wrinkling \cite{zhang2013adhesion}. Nonetheless, we point out that the frictional forces between graphene and hBN will still affect the wrinkle pattern, for example influencing the separation between different wrinkles \cite{zhang2013adhesion, zhang2014understanding}. Moreover, we argue that interfacial friction will make each wrinkle more mechanically stable by opposing to successive relaxations of the flake, which would result in the collapse of the wrinkle. The device structure is schematically sketched in Fig.~\ref{fig:characterization}(a). An optical image of a representative device is displayed in Fig.~\ref{fig:characterization}(b). The hBN thickness in this case is 40 nm.

Before transferring graphene onto hBN, we acquired a Raman map of the flake deposited on PMMA (see Appendix \ref{app:fab}). Since the 2D peak fitted into a single Lorentzian, we concluded that the flake was monolayer \cite{ferrari2007raman, hao2010probing}. Then, we deposited the graphene flake partially on hBN and partially on SiO$_2$ acquiring Raman spectra on both sectors. The spectra in Fig. \ref{fig:characterization}(c) highlight the difference between graphene deposited on SiO$_2$ (left panel) and on hBN (right panel). Measurements were performed with a 532 nm solid-state laser via a spatially resolved micro-Raman setup ({\it inVia Renishaw}) using a 100x objective. We observe that the spectra acquired on graphene placed on top of hBN are characterized by a very strong background due to hBN luminescence. Nonetheless, the D peak is negligible in all spectra and thus we infer that the exfoliated flake lacks significant defective regions. Moreover, the 2D peak is symmetric and can be fitted into a single Lorentzian whose $\Gamma_{\rm 2D}$ is typically of the order of 25 cm$^{-1}$ for both substrates, SiO$_2$ and hBN.

We etched the flakes to obtain a set of standardized $10\times10$ $ \mu$m$^2$ graphene squares on hBN, as described in Appendix \ref{app:fab}, and characterized their morphology using a commercial {\it Bruker's Dimension Icon} Atomic Force Microscope (AFM) with a {\it ScanAssyst-Air} probe operating in Peak Force mode. We remark that the flake displayed in Fig.~\ref{fig:characterization}(d), representative of all the flakes under study, is highly inhomogeneous with large polygonal bubbles which have approximate lateral size of the order of 1 $\mu$m and height between 50 and 100 nm. These bubbles are distributed on the whole graphene surface. Wrinkles with nanometric height and $\sim 10$ nm width originate from their vertices \cite{zhang2017coexistence}. Interestingly, the bubbles correspond to a strong redshift of the 2D mode peak frequency ($\omega_{\rm 2D}$) of Fig.~\ref{fig:characterization}(e) as well as to its broadening (not shown): $\Gamma_{\rm 2D}$ reaches values up to 70 cm$^{-1}$. Since we do not expect a significant change of the doping in correspondence of the bubbles, we argue that these effects are mainly due to a local tensile strain on the flake \cite{neumann2015raman}.

\section{Deformation using polymeric actuators} \label{sec:crumpling}
\begin{figure}[h!]
    \centering
    \includegraphics[scale=0.25]{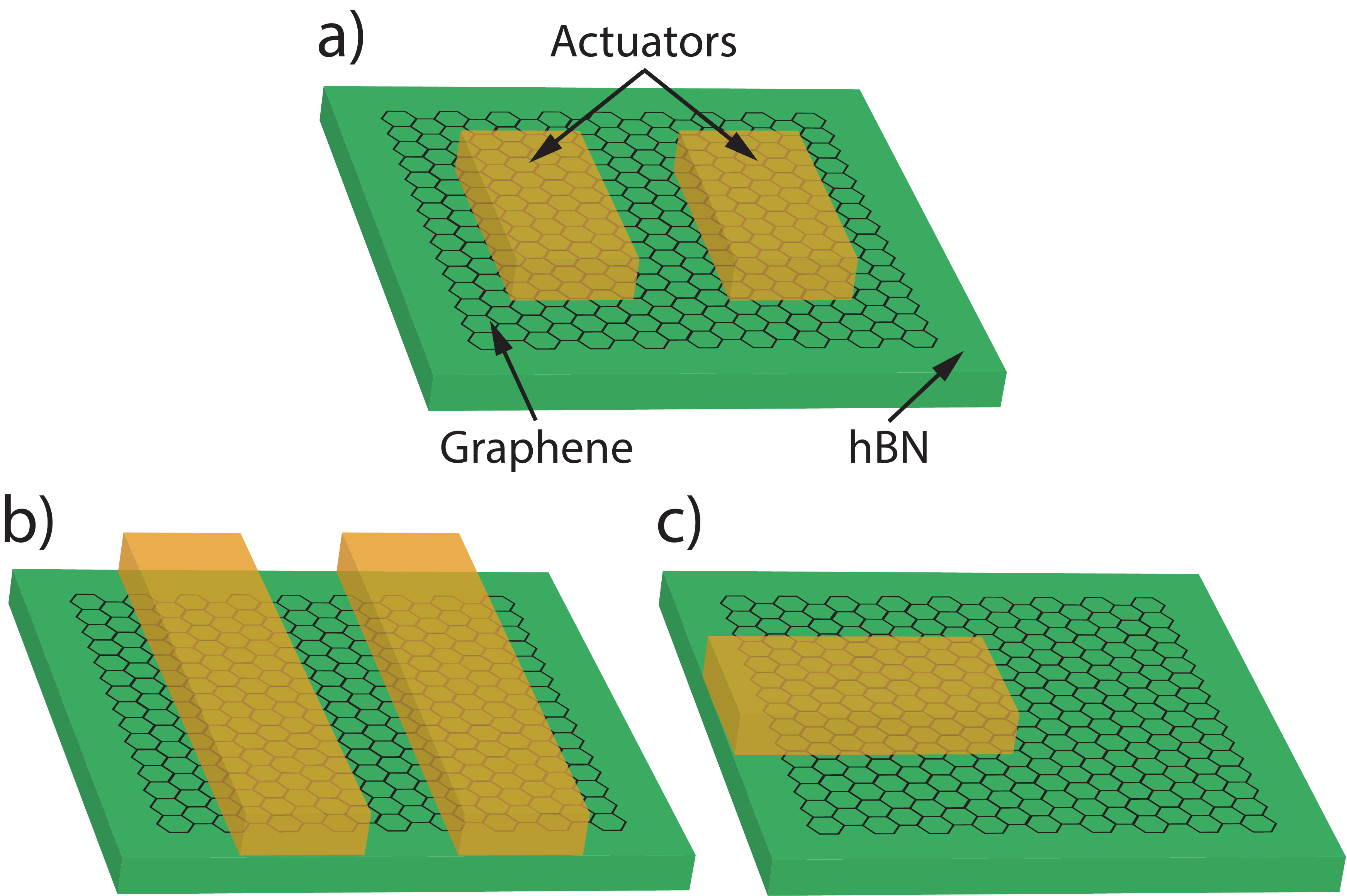}
    \caption{Sketch of the actuators geometries. a) A pair of parallel un-anchored actuators (type-A geometry). b) A pair of parallel actuators anchored at both ends (type-B). c) A single actuator anchored at one end (type-C).}
    \label{fig:geometries}
\end{figure}
We tested the deterministic wrinkling on 5 devices with 3 different actuator geometries (see Fig. \ref{fig:geometries} and Appendix \ref{app:fab} for details on the fabrication process). The first device consisted of two parallel PMMA actuators placed in the middle of the flake as schematically sketched in Fig.~\ref{fig:geometries}(a). We label this device geometry as type-A. Fig.~\ref{fig:geometries}(b) shows a second actuator geometry: type-B devices consisted of two long parallel PMMA stripes crossing vertically the whole flake with their final portion overlapping the bare hBN. As a result, PMMA anchored the edges of the flake to the underlying hBN, thus suppressing vertical contraction. The last actuator geometry (type-C) is displayed in Fig.~\ref{fig:geometries}(c) and consisted of a single PMMA stripe which was anchored to the hBN substrate on one of its ends. In general, we expect that the PMMA actuators, interacting with the underlying graphene through Van der Waals forces \cite{vivas2022chemical}, exert a compressive force which pulls and deforms graphene thanks to the low friction between graphene and hBN \cite{leven2013robust,wang2016thermally}. The resulting squeezing induces out-of-plane deformations which collapse into isolated wrinkles and depends on the shape of the actuators and of the flake. The actuation of the PMMA stripes was performed by iteratively scanning PMMA with an electron-beam at 2 keV resulting in the shrinkage of the PMMA patches. More details on this technique can be found in ref. \cite{giambastiani2020stress}.

\begin{figure}
    \centering
    \includegraphics[scale=0.1]{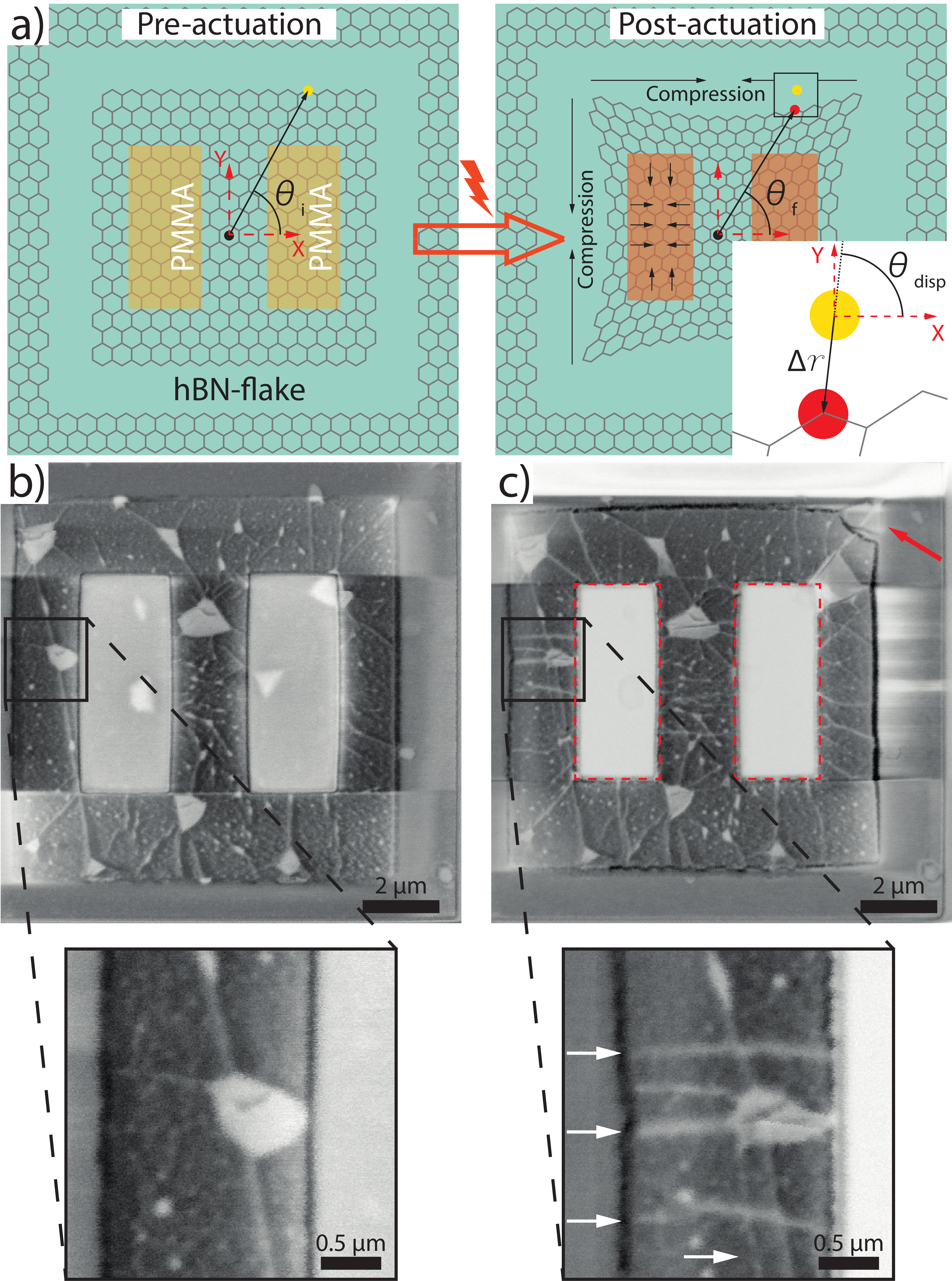}
    \caption{a) Sketch of the device before (left) and after (right) the electron irradiation of the PMMA actuators. Inset panel: magnified view of the area inside the black square. b) SEM image of one of the studied devices before the actuation. Inset panel: magnified view of the area inside the black square. c) SEM image of the same device after the actuation. The dashed red squares highlight the area of electron irradiation. The red arrow highlights a small triangular flake region torn off from the main one. The scales of the images are the same as those in panel (b). Magnified view (black square) of a region with induced wrinkles. Inset panel: magnified view of the area inside the black square where the white arrows highlight wrinkles induced via electron irradiation.}
    \label{fig:sketch}
\end{figure}

We schematically sketch the expected deformation of type-A device in Fig.~\ref{fig:sketch}(a). Comparing the pre- and post-actuation Scanning-Electron Microscope images, respectively displayed in Fig.~\ref{fig:sketch}(b) and (c), we observe a deformation of the edges which suggests an accompanying change in the morphology of the flake. The deformation effect is particularly evident when looking at the top-right vertex where a triangular portion of the flake was torn off from the main one. A widespread wrinkling of graphene can be clearly observed both between the two PMMA actuators and between each actuators and the flake edges. Note that wrinkles formation tends to occur along preferential directions which are perpendicular to the edges of the actuators and typically originating from the edges of the flake.

Additional wrinkles are correlated to the deformation of the pre-actuation bubbles. This latter deformation process could be also due to the heating of the trapped gases at the graphene/hBN interface \cite{haigh2012cross,leconte2017graphene} which can expand due to electron-beam induced heating and deform the bubbles. This, together with the mechanical action of the polymeric actuators, promotes the formation of a significant number of new wrinkle structures.
\begin{figure*}
    \centering
    \includegraphics[scale=0.3]{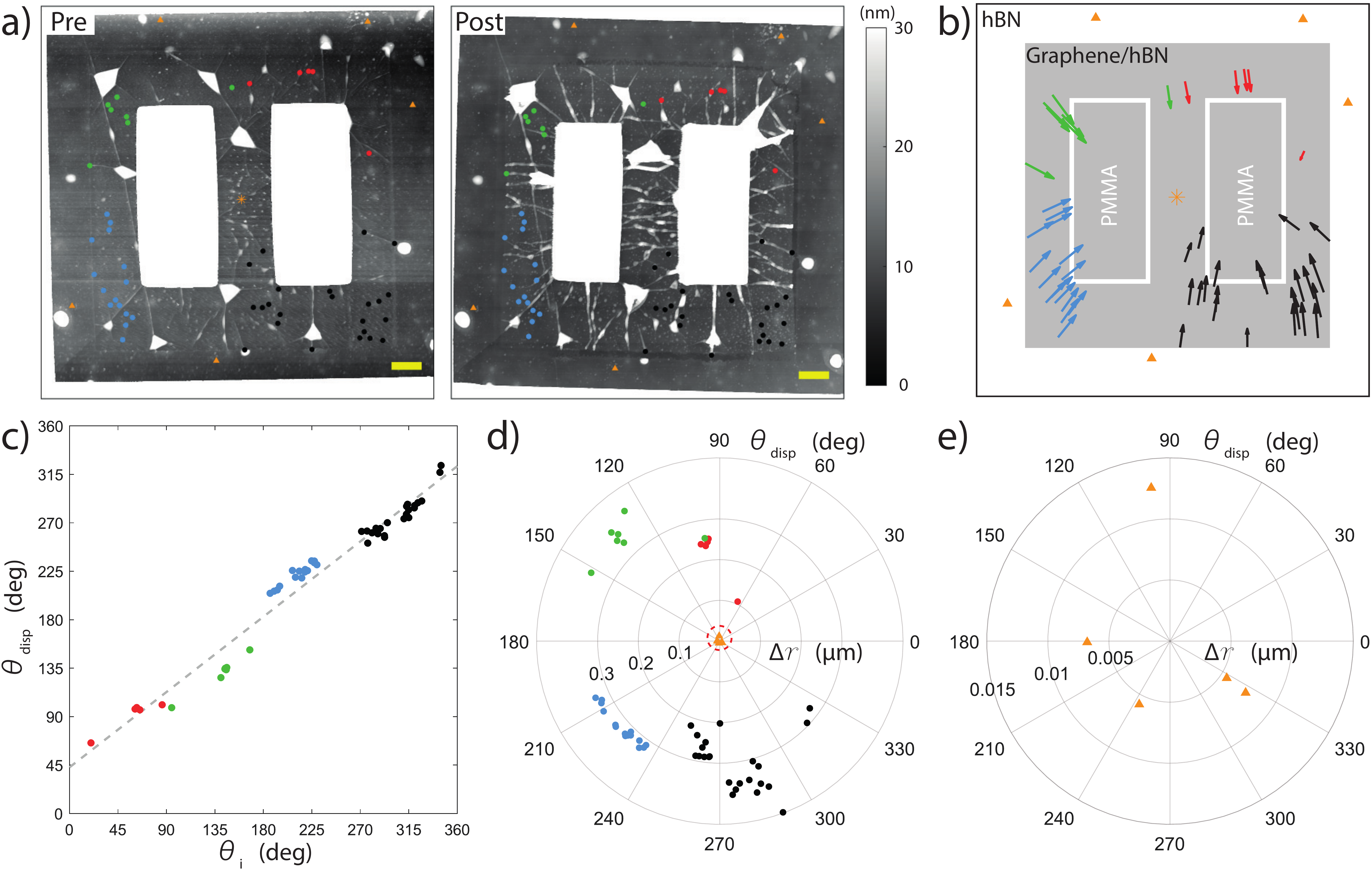}
    \caption{a) AFM scan of the device of Fig. \ref{fig:sketch} before (left) and after (right) the electron irradiation of the PMMA. Different colors are used to identify reference markers in qualitatively different regions of the device. The yellow scalebar is 1 $\mu$m. b) Top-view sketch of the device. The coloured arrows represent the direction of the displacement, with their length corresponding to the physical displacement magnified by a factor 3. c) Angular displacement of the markers displayed in panel (a) with the same color coding. d) Polar plot displaying the magnitude of the displacement of the GRMs ($\Delta r$) vs. the direction of the displacement ($\theta_{\rm disp}$). e) Magnified view of the polar plot displayed in panel (d) (see the red dashed circle) highlighting the displacement of the SGMs.}
    \label{fig:AFM1}
\end{figure*}

We argue that wrinkling is not only determined by the action of the actuators or by the heating of entrapped gases, but also by the deformation of the edges which promotes the detachment of graphene from the substrate and the creation of wrinkles propagating on the entire flake. Thus, it is pivotal to map the squeezing pattern of graphene in order to understand the respective role of the actuator and flake geometries.

In order to precisely measure the effects of the electron irradiation on the actuators, we analyzed the AFM maps of each flake. Operationally, we acquired an AFM-scan before and after each irradiation step. To compare the corresponding pre- and post-actuation scans, we picked some fixed reference points on the hBN flake (later referred to as substrate global markers; SGMs). Then, we aligned the two images through a roto-translation operation minimizing the SGMs post-actuation displacement. The evaluation of the graphene squeezing was performed by measuring the displacement of the same set of selected graphene reference markers (GRMs) before and after actuation. These markers were manually picked by choosing recognizable features on the graphene flake, such as polymeric residues left by previous fabrication processes. The deformation of graphene was quantified by measuring the position of the same markers before and after actuation. A clarifying example of our method is reported in the sketch in Fig.~\ref{fig:sketch}(a). After the alignment procedure, we started by defining a GRM in the pre-actuation map (yellow dot). After electron irradiation, the graphene flake was squeezed, and the GRM was displaced to a new position (red dot). Before actuation (left panel), the position of each GRM was identified by the initial angle $\theta_{\rm i}$. After the actuation (right panel), the position of the marker changed, due to the compression of the flake, and its position was identified by the angle $\theta_{\rm f}$. These two angles, $\theta_{\rm i}$ and $\theta_{\rm f}$, were defined in the same coordinate system and were measured with respect to the X axis. In order to better visualize the final position of the GRM, we use the displacement angle $\theta_{\rm disp}$, defined in the inset of Fig.~\ref{fig:sketch}(a), in place of $\theta_{\rm f}$. We point out that $\theta_{\rm disp}$ is measured in the marker's reference system. The displacement magnitude of each GRM is given by $\Delta r$, instead. We use the same procedure to show the displacement of the SGMs. The common origin for these coordinates corresponds to the intersection of the diagonals linking the vertices of each flake which are identified manually in the pre-actuation image, then reported to the post-actuation one. Using this coordinate system, we can proceed to a quantitative analysis of the buckling geometry.
\begin{figure*}
    \centering
    \includegraphics[scale=0.25]{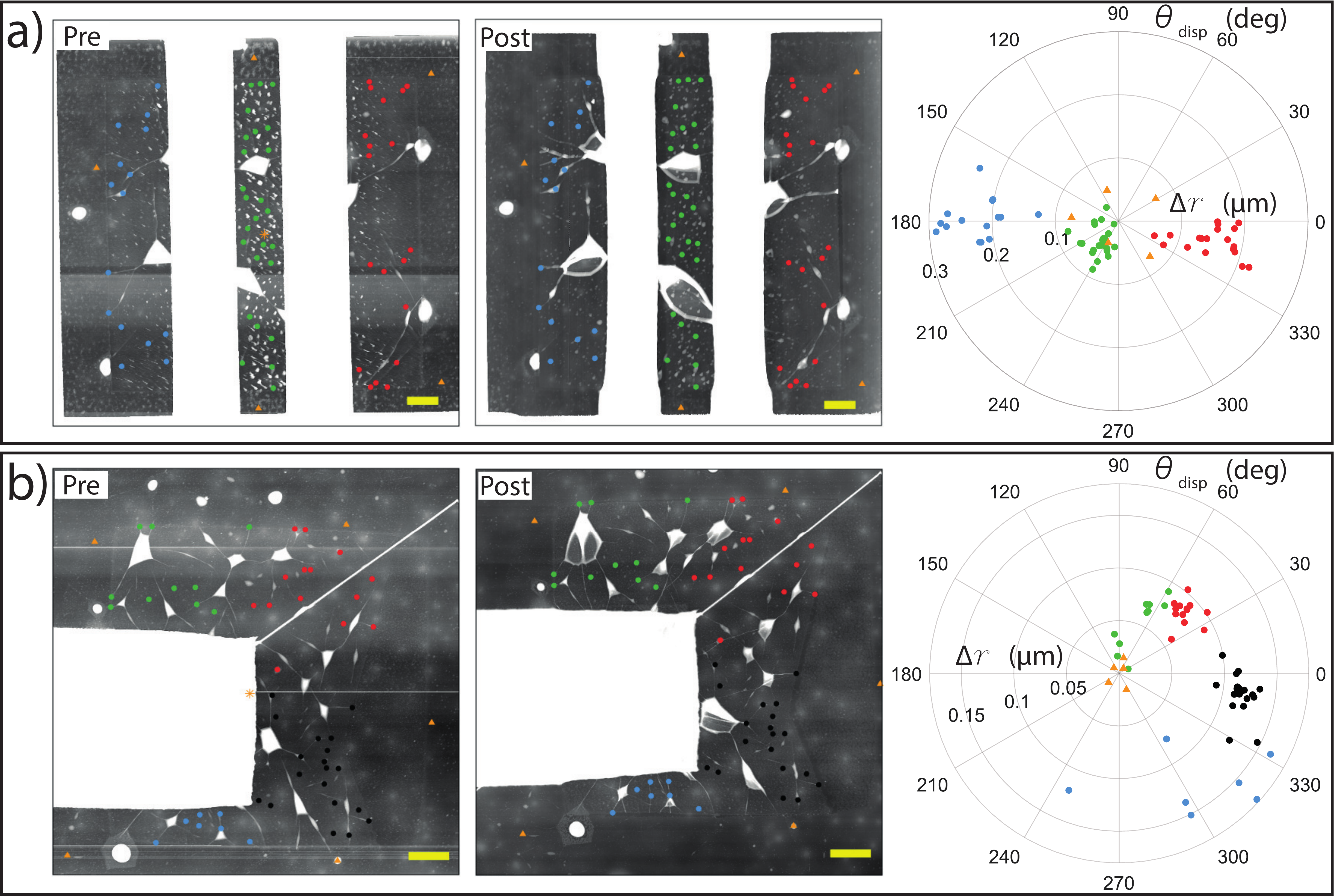}
    \caption{Buckling in alternative actuator geometries. a) From left to right: pre-, post-actuation AFM scan and polar plot of the displacement of the SGMs and of the GRMs for a type-B device. b) Equivalent study of panel a for type-C device. The yellow scalebars all correspond to 1 $\mu$m and the vertical scalebar is the same displayed in Fig.~\ref{fig:AFM1}(a).}
    \label{fig:AFM2}
\end{figure*}

The AFM scans are reported in Fig.~\ref{fig:AFM1}(a) where the marker's colors (red, black, blue and green) identify the location of the markers inside 4 possible quadrants (top-right, bottom-right, bottom-left and top-left, respectively). After electron irradiation of the actuators (right panel), we observe a pervasive wrinkling of the graphene surface inducing several localized wrinkles. We note that wrinkles tend to nucleate perpendicularly to the edges of the flake as well as from the deforming bubbles. Similarly, we observe the formation of a set of parallel wrinkles in the central area between the two actuators. We also highlight that the sliding and deformation of the flake is crucial for the development of such structures. In fact, the PMMA actuation clearly shifts the GRMs towards the center and distorts the edges of the flake. The graphene compression can be visualized by plotting the displacement vector (magnified by a factor $\times3$, see Fig.~\ref{fig:AFM1}(b)). We remark that the magnitude of the displacement has a roughly circular symmetry. The displacement direction of the GRMs is displayed in Fig.~\ref{fig:AFM1}(c): this approximately follows $\theta_{\rm i} \sim \theta_{\rm disp}$, which again hints at an average circularly-symmetric contraction. From Fig.~\ref{fig:AFM1}(d), we note a very good alignment of the pre- and post-AFM scans by noticing that the displacement of the SGMs (orange triangles) is negligible, being roughly a ten of nanometers (see Fig.~\ref{fig:AFM1}(e)), while the typical displacement of the GRMs is of the order of hundreds of nanometers. The phenomenology observed for this device was also found in the two type-A devices shown in Appendix \ref{app:additional}.

The type of graphene squeezing can be controlled by changing the buckling geometry. In Fig.~\ref{fig:AFM2}(a), we display the deformation pattern produced by type-B geometry. We point out that type-A device, lacking any constrain, is squeezed both in the vertical and horizontal directions. In the case of type-B device, the PMMA cannot freely contract along the vertical direction. Thus, in analogy with type-A device, we would expect the actuation of type-B device to induce a quasi-uniaxial deformation of graphene in the unconstrained direction (perpendicularly to the two PMMA stripes). From the pre- and post-irradiation AFM scans displayed in Fig.~\ref{fig:AFM2}(a), we observe that the PMMA actuation is much less efficient than type-A geometry in generating deterministic arrays of wrinkles: wrinkles indeed originate mostly from the deformation of bubbles. In this case, we dissect the flake in three different regions identified by three different colors: the two side-regions (left and right) and the central region (between the two vertical stripes). The displacement within the latter region (green dots) is vanishing within the alignment uncertainty (orange triangles): this is expected from the left-right mirror symmetry of the device. The blue and red GRMs, which correspond to left and right regions respectively, show instead a displacement ranging between 200 and 300 nm. Moreover, the displacement is directed almost horizontally towards the center of the flake, as clearly visible in Fig.~\ref{fig:AFM2}(a), supporting the hypothesis of roughly uniaxial deformations.

The effect of electron irradiation on Type-C device is displayed in Fig.~\ref{fig:AFM2}(c). This actuator was designed to laterally pull the underlying flake and induce a roto-translation. We stress that the development of this kind of actuator could also permit precise sliding and alignment of graphene on hBN, ultimately leading to physically relevant twisted bilayer heterostructures \cite{du2020moire,de2018strain,yang2020situ}. Analogously to the previous type-B case, we report a strong modification of the bubbles with poor wrinkle formation. The GRMs mainly undergo a leftward displacement: globally, the magnitude of the displacement of the GRMs is maximized around $\theta_{\rm disp} = 0^{\circ}$ and tends to decrease approaching the left edge of graphene (see the polar plot in Fig.~\ref{fig:AFM2}(c)). This behaviour is generally expected since the flake is anchored to its left edge, but the exact sliding geometry is complicated by the presence of inhomogeneities of the substrate.

\section{Simulations} \label{sec:sim}

In order to interpret the wrinkling effect observed in the previous sections, we simulated the mechanical behaviour of the device using a finite-element solver (COMSOL Multiphysics). We modeled a $10\times10$ $\mu$m$^2$ continuous and isotropic flake of graphene with Young's modulus of 1 TPa and Poisson's ratio of 0.15 \cite{kudin2001}. The squeezing was modeled by introducing a planar stress in the area of graphene covered by the PMMA pads, assuming that PMMA and graphene cannot slide one with respect to the other. In this way, stress is transferred from the polymer to the graphene, according to their joint mechanical response. Further important approximations were made. First, the flake deformation was modelled in a two-dimensional space; thus, we assess the possibility of wrinkle formation based on the presence of compressive stresses, but we neglected out-of-plane buckling of the flake and did not take into account the impact of adhesive forces at the graphene/hBN interface. Second, we neglected the frictional interactions between graphene and hBN. These two approximations could strongly affect the behaviour and the geometrical parameters of the resulting wrinkle pattern whose formation could also be suppressed by strong adhesion or friction at the interface \cite{zhang2013adhesion, zhang2014understanding}. Moreover, by assuming a continuous and isotropic flake of graphene we also neglected the crystal structure of the flake whose six-fold symmetry could affect the exact observed morphology and deformation pattern.

\begin{figure}
    \centering
    \includegraphics[scale=0.3]{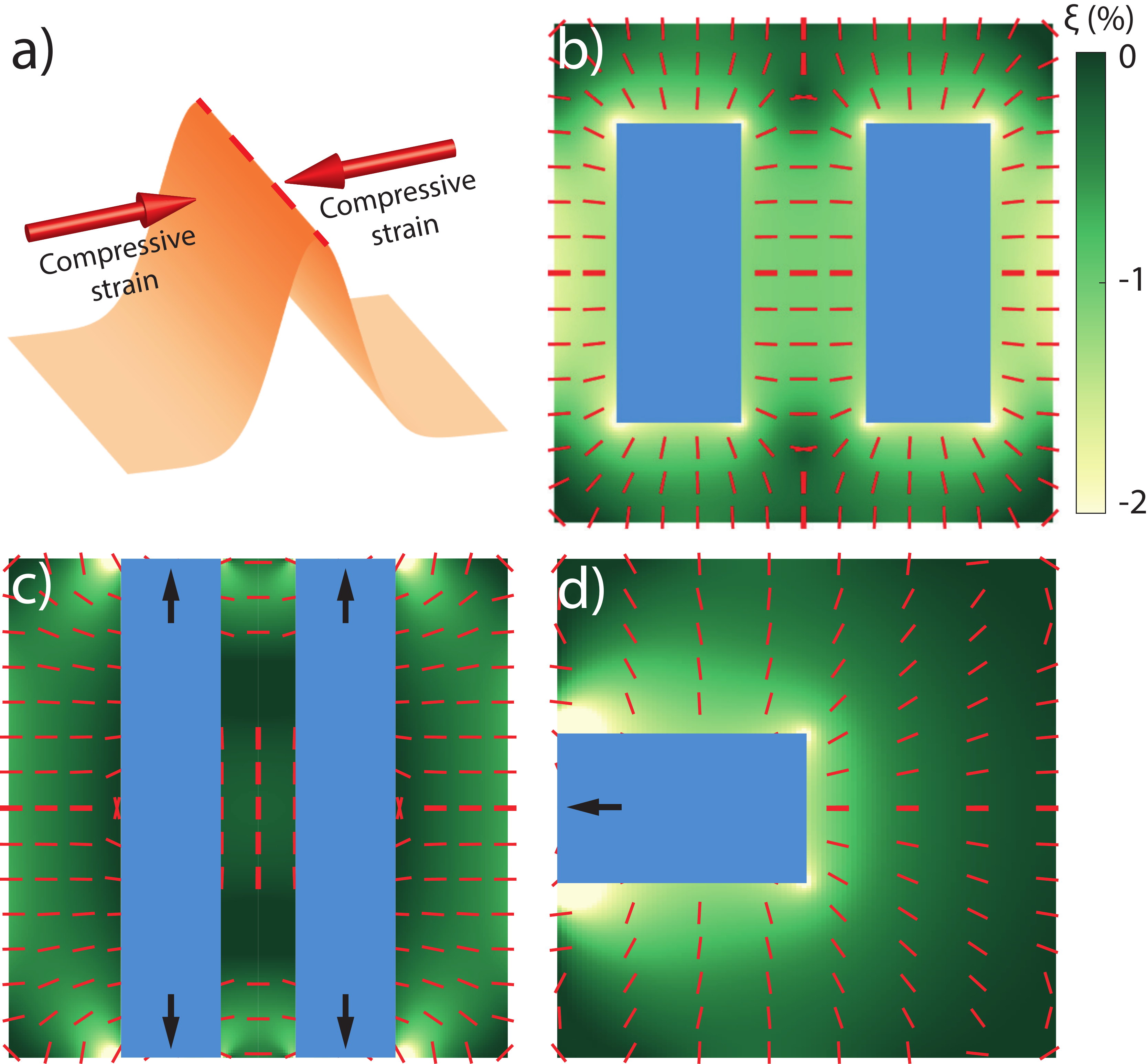}
    \caption{Numerical simulations. a) Schematic sketch of the formation mechanism of a wrinkle. b) Colorplot of the smallest negative eigenvalue obtained by diagonalizing the strain tensor. The blue patches represent the PMMA actuators for a type-A geometry. The red lines depict the expected wrinkle direction. c) Same sketch in panel (b) for type-B geometry. The black arrows indicate the anchored graphene edge. d) Same sketch in panel (b) for type-C. The black arrow indicates the anchored graphene edge.}
    \label{fig:sim}
\end{figure}

The direction of the wrinkles can be inferred by the direction of the eigenvectors of the system's strain tensor. Since wrinkles are physically induced by a compressive force acting on graphene, we expect that their direction is orthogonal to the negative eigenvector, corresponding to a compressive deformation, as displayed in Fig.~\ref{fig:sim}(a). The magnitude of the compressive (negative) eigenvalue is displayed in the colorplot of Fig.~\ref{fig:sim}(b), where the red lines represent the expected wrinkle direction.

We highlight that the calculated preferential direction for wrinkles in type-A geometry is perpendicular to the two PMMA actuators, as observed in the actual irradiation experiment described in section \ref{sec:crumpling}. We also highlight that the formation of wrinkles is favoured in regions where the magnitude of the local compressive strain is larger. For the simulation of type-B and -C devices we modeled the constrained graphene edges by imposing the condition of zero-displacement at the boundaries indicated by the black arrows in figure \ref{fig:sim}(c) and (d). For type-B geometry (see Fig.~\ref{fig:sim}(c)), we expect wrinkles to be mainly perpendicular to the PMMA actuators: this agrees well with what has been found for type-A device since the two kinds of devices are relatively similar. Nonetheless, we argue that, in this case, the PMMA anchors the flake, which can only deform quasi-uniaxially. This kind of constrained deformation thus turns out to be inefficient in creating isolated wrinkles due to the suppression of the compressive eigenvalue (see figure \ref{fig:sim}(c)). Similarly, simulations indicate that type-C device (see Fig.~\ref{fig:sim}(d)) should induce a pattern of wrinkles perpendicular to the actuator edges. In our experiment, we did not observe the formation of wrinkles (see figure \ref{fig:AFM2}b). This was probably due to the anchoring effect of the PMMA actuator, suppressing the compressive strain, and the roughness of the hBN surface which did not allow free sliding and wrinkling.

\section{Conclusion} \label{sec:conclusion}

In this paper, we induced the deformation of exfoliated flakes of graphene deposited onto a thick and flat hBN flake by using polymeric actuators patterned on top of graphene. We identify one promising actuator geometry to induce a relevant and controllable wrinkling patterns caused by an overall squeezing of the flake. Since hBN represents one of the best insulating materials for graphene-based devices \cite{leconte2017graphene}, the possibility to tune the appearance of deterministic wrinkles on graphene/hBN heterostructures offers a promising platform to study and exploit the electronic effects of wrinkles in graphene. Moreover, polymeric micro-actuators \cite{colangelo2018controlling,colangelo2019local,gilbert2019strain} could be used not only to induce wrinkles by folding the graphene sheet but also to create and release complex wrinkle architectures via folding-unfolding cycles thanks to properly designed geometries. These could also be used to control the geometric distribution of wrinkles and design origami-based devices \cite{zhao2019geometry} or annular geometries which could be exploited for applications in nano-force sensors and tunable magnetic and electronic devices \cite{li2015surface}. Finally, the graphene-based devices shown in this paper represent a proof of concept which, in principle, can be extended to arbitrary two-dimensional materials.

\section{Acknowledgments}

DG, CT and SR thank S. Cassandra for helping with the preparation of the key image. DG thanks G. Ciampalini for useful discussions, and A. Guerrini and A. Moscardini for the technical support. We acknowledge the financial support of the Italian Ministry of University and Research (PRIN Project QUANTUM2D and PRIN Project MONSTRE-2D). The research leading to these results has received funding from the European Union’s Horizon 2020 research and innovation program under grant agreement no. 881603-Graphene Core3.

\appendix

\section{Device fabrication} \label{app:fab}
\subsection{Mechanical exfoliation}

We exfoliated the hBN flakes by iteratively applying an adhesive tape to bulk boron nitride and stuck the tape on a n-type doped silicon chip coated with 300 nm of SiO$_2$ and resistivity of 1-5 m$\Omega \cdot$cm. The substrate was previously cleaned in acetone for 10 minutes and in O$_2$-plasma at 100 W for 10 minutes. After cleaning in acetone, the chip was rinsed in isopropyl alcohol.

\begin{figure}[]
    \centering
    \includegraphics[scale=0.25]{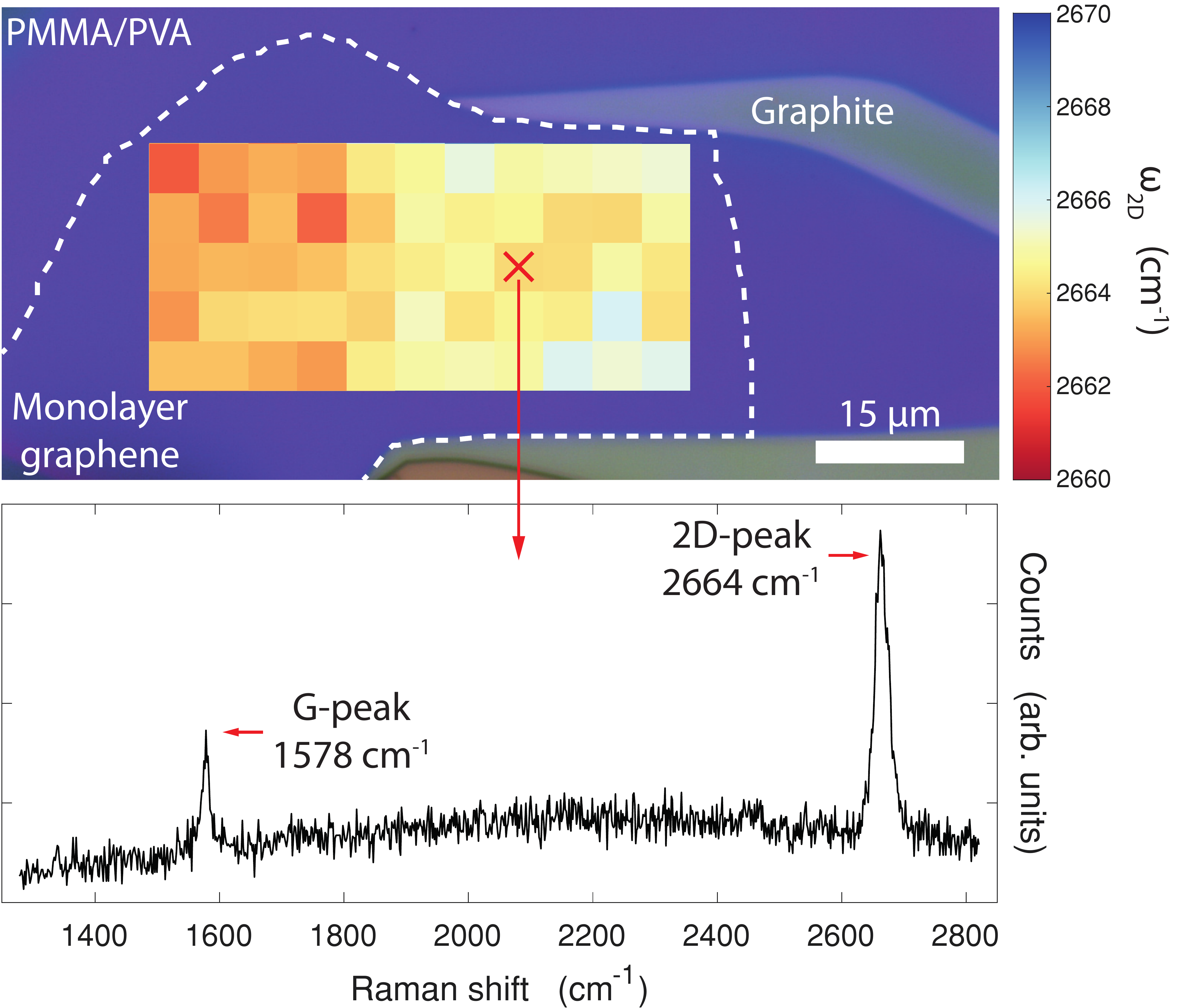}
    \caption{Raman map of the 2D peak position superposed to an optical image of the flake in Fig. \ref{fig:characterization} onto the PMMA/PVA substrate (before the transfer process). The white dashed line highlights the edge of the monolayer flake.}
    \label{figS4}
\end{figure}

\begin{figure*}[]
    \centering
    \includegraphics[scale=0.45]{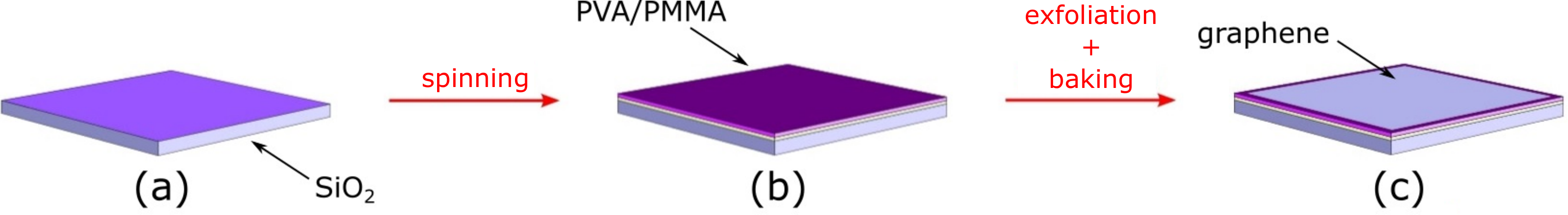}
    \caption{Schematic of the exfoliation process for graphene. A SiO$_2$ chip (a) is spin-coated with a PVA layer and then a PMMA layer (b); graphene is exfoliated using adhesive tape, which is then attached onto the PMMA layer; the chip is baked and then the tape is peeled off, leaving graphene on PMMA (c).}
    \label{figS0}
\end{figure*}

In order to control the placement of selected monolayer graphene flakes on the target hBN flake, we performed a preliminary exfoliation step on a polymeric layer, which allowed for a more accurate transfer process. We prepared a large silicon chip (about $2\times2$ cm)  by spin-coating a layer of polyvinyl alcohol (PVA), at 4000 rpm for 1 minute, followed by a layer of PMMA (AR-679.04), at 2800 rpm for 1 minute. Each spin-coating step was followed by a bake at $150\,^{\circ}$C for 1 minute. After the preparation of the substrate, we iteratively exfoliated bulk graphite (this procedure was iterated from 2 to 5 times) using adhesive tape. We then stuck the tape to the PMMA layer and baked it for 5 minutes at $170\,^{\circ}$C. The last step consisted in slowly detaching the tape from the PMMA layer. In figure \ref{figS4}, we display one flake of monolayer graphene and the corresponding 2D peak position ($\omega_{\rm 2D}$) before the transfer process. The Raman spectra are acquired with a 50x objective at 0.5 mW. The whole transfer process is sketched in Fig. \ref{figS0}.

\subsection{Transfer and reactive-ion etching of graphene}

Before transferring the graphene flake on the selected hBN flake, we chemically cleaned hBN in acetone and, later, in AR 600-71 (rinsed in H$_2$O) for 10 minutes. Moreover, in order to reduce the formation of bubbles at the graphene-hBN interface, we also annealed the chip with the hBN flakes in N$_2$ atmosphere at $500 \,^{\circ}$C for 15 minutes and cleaned the sample in O$_2$-plasma at 30 W for 10 seconds.

To transfer the graphene flake on the selected hBN flake, we dissolved the PVA layer in H$_2$O at about $50\,^{\circ}$C, leaving a floating PMMA membrane with graphene on top. We picked up the PMMA membrane and transferred the graphene flake on the hBN flake. During the transfer, the membrane was slowly approached to the target chip heated at $105\,^{\circ}$C. After the transfer procedure, the sample was cleaned in acetone and in AR 600-71 for 2-5 hours in order to dissolve any polymeric residue. Then, we spin-coated a layer of PMMA (AR-679.02) at 2000 rpm and baked at $90\,^{\circ}$C for 15 minutes. We patterned PMMA squares which were used as polymeric masks to etch graphene in O$_2$-plasma for 30 seconds.

\subsection{PMMA actuators}

After one last cleaning step in acetone and AR 600-71, we spin-coated PMMA (AR-P~679.02) on the sample at 2000 rpm and baked it at $90^{\circ}$C for 15 minutes. We patterned the PMMA actuators via electron-beam lithography (20 keV electrons with an electron dose of 280 $\mu$C/cm$^2$), in correspondence of the square graphene flakes defined by the etching process.

\section{Numerical simulation}

\begin{figure}[h!]
    \centering
    \includegraphics[scale=0.65]{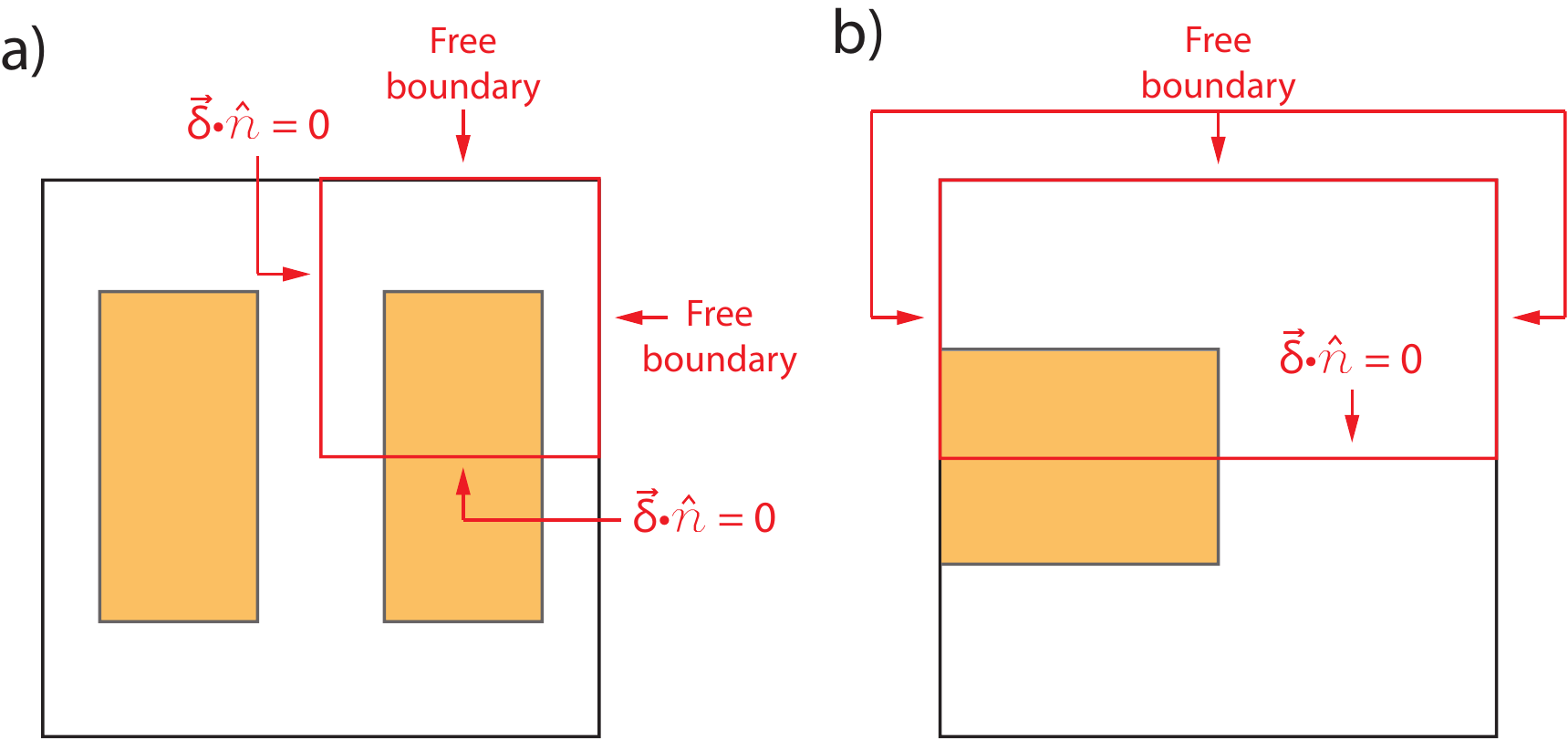}
    \caption{a) Primitive cell (red edges) and boundary conditions applied for the simulation of type-A device (the same applied for the type-B device). b) Primitive cell (red edges) and boundary conditions applied for the simulation of type-C device. The orange rectangles correspond to the PMMA actuators.}
    \label{figS3}
\end{figure}
As discussed in the main manuscript, we numerically simulated the devices to calculate the strain tensor and predict the wrinkle geometry. For the numerical simulation of the mechanical problem, we split the device geometries in sub-cells, depending on their symmetries. We calculated the solution inside just one of the sub-cells (primitive cell) by imposing the proper boundary conditions. In order to get the final result, we mirrored the primitive cell to recover the full geometry. Type-A and B have a four-fold mirror symmetry and we restricted our simulation to the top-right primitive cell, indicated with red edges in Fig. \ref{figS3}(a). The internal edges of the primitive cell were subjected only to displacements $\vec{\delta}$ parallel to the edge, i.e. $\vec{\delta}\cdot\hat{n}=0$, where $\hat{n}$ is the unitary vector normal to the edge. The external edges of the flake were free to deform apart from the boundaries indicated by the black arrows in figure \ref{fig:sim}(c). Type-C device has a two-fold mirror symmetry. One of the two rectangles can be taken as primitive cell, see red edges in Fig. \ref{figS3}(b). The boundary conditions are displayed in Fig. \ref{figS3}(b). The external edges of the flake were free to deform apart from the boundary indicated by the black arrow in figure \ref{fig:sim}(d).

\section{Additional datasets} \label{app:additional}
\begin{figure}[h!]
    \centering
    \includegraphics[scale=0.3]{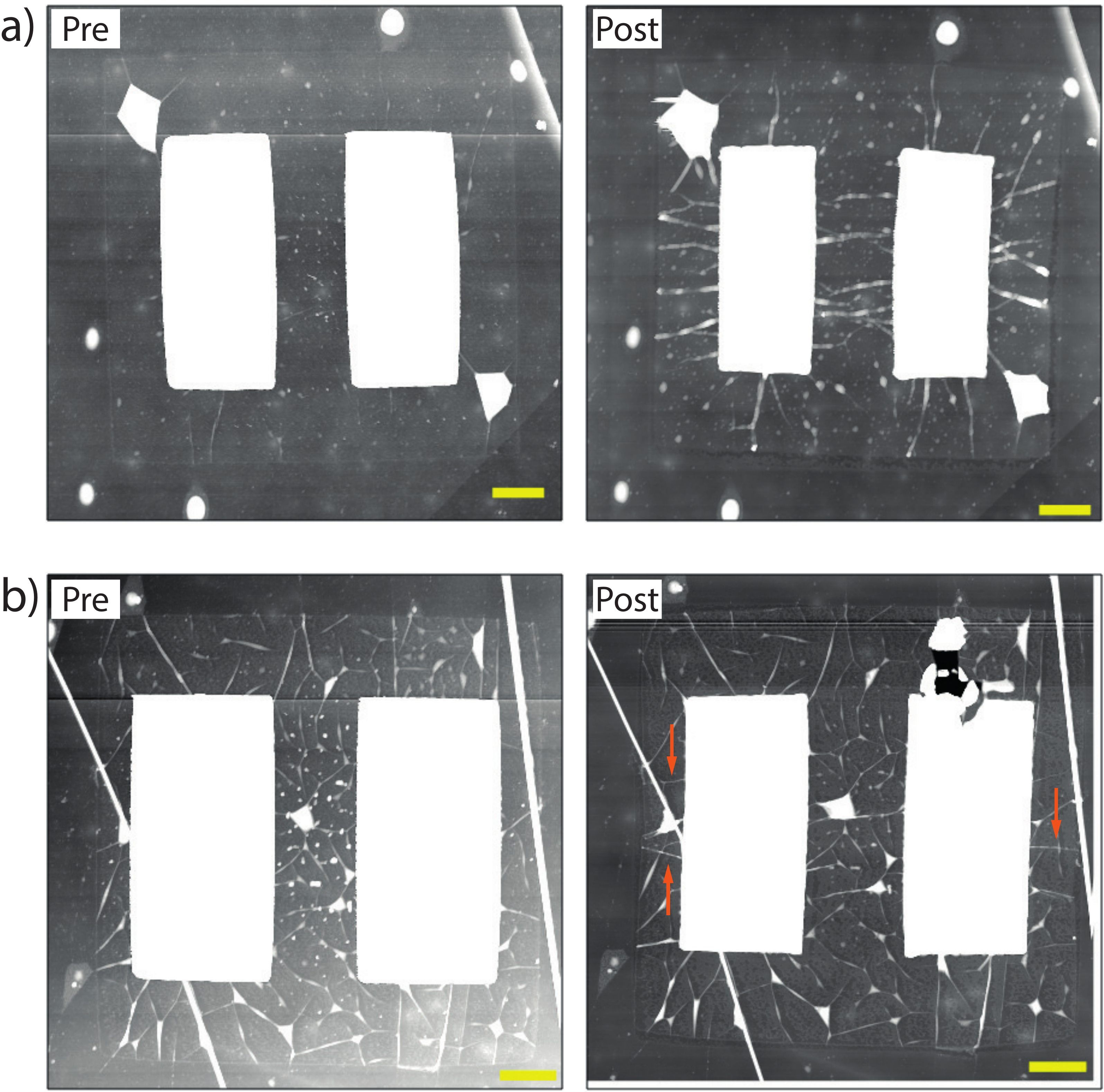}
    \caption{AFM-scans of two type-A devices (panels (a) and (b)) before (left) and after (right) electron irradiation. The orange arrows in the post-irradiation scan of panel (b) highlight the presence of some wrinkles which were not present before the irradiation. The yellow scalebars inside the AFM scans are 1 $\mu$m and the AFM colorbar is the same displayed in Fig.~\ref{fig:AFM1}(a).}
    \label{figS1}
\end{figure}

We report the results for two additional type-A devices in Fig. \ref{figS1}(a) and (b). The main dissimilarity between the two devices is the different number of wrinkles which were present before the electron irradiation. In particular, the pre-actuation scan displayed in panel (b) shows a flake which is much more wrinkled than the one in panel (a). This is probably correlated to the fact that the fabrication of the device displayed in panel (b) did not include any annealing in N$_2$ atmosphere nor cleaning in O$_2$ plasma of the hBN flake before graphene transfer. We argue that these steps are pivotal to flatten and improve the homogeneity of the hBN flake and promote the sliding of the transferred graphene.

The different behaviour of the two flakes becomes apparent in the two AFM-scans acquired after electron irradiation. The device displayed in panel (a) shows a high degree of wrinkle formation, similarly to the flake displayed in Fig. \ref{fig:AFM1}(a). On the contrary, the device in panel (b) barely features the formation of 3 new wrinkles, highlighted by the orange arrows. We note that all the 3 wrinkles originate at the edges of the flake. This remarks the importance of edges which are free to deform in the formation of the wrinkle pattern. Moreover, we argue that this confirms the importance of the hBN flatness and homogeneity to induce the formation of wrinkles. However, we observe that the absence of wrinkles in this device could also be due to the six-fold symmetry of graphene which could alter the membrane response to strain in different crystal directions.

\bibliographystyle{BSTart}
\bibliography{Refs}

\end{document}